\documentclass{jfm}
\usepackage{color}
\usepackage{amstext}
\usepackage{amssymb}
\usepackage{graphicx}
\usepackage{amsmath}

\newcommand\const{\mathrm{const}}

\newcommand\vA{\boldsymbol{A}}

\newcommand\vO{\boldsymbol{O}}

\newcommand\vF{\boldsymbol{F}}

\newcommand\va{\boldsymbol{a}}

\newcommand\vu{\boldsymbol{u}}
\newcommand\vv{\boldsymbol{v}}

\newcommand\vx{\boldsymbol{x}}

\newcommand\vg{\boldsymbol{g}}

\begin{document}

{\title[Many Faces of Boussinesq Approximations ] {Many Faces of Boussinesq Approximations}}

\author[V. Vladimirov and N. Al-Salti]
{V.\ns A.\ns V\ls l\ls a\ls d\ls i\ls m\ls i\ls r\ls o\ls v\ns and\ns N.\ns A\ls l\ls \ls -\ls S\ls a\ls l\ls t\ls i\ls}

\affiliation{Sultan Qaboos University, Oman, University of York and University of Cambridge, UK}

\date{November 12th 2016}

\setcounter{page}{1}\maketitle \thispagestyle{empty}

\begin{abstract}

 The \emph{equations of Boussinesq approximation} (EBA) for an incompressible and inhomogeneous in density  fluid are analyzed from a viewpoint of the asymptotic theory.
 A systematic scaling shows that there is an infinite number of related asymptotic models. We have divided them into three classes: `poor', `reasonable' and `good' Boussinesq approximations. Each model can be characterized by two parameters $q$ and $k$, where $q =1, 2, 3, \dots$ and $k=0, \pm 1, \pm 2,\dots$. Parameter $q$ is related to  the `quality' of approximation, while $k$ gives us an infinite set of possible scales of velocity, time, viscosity, \emph{etc.} Increasing  $q$ improves the quality of a model, but narrows the limits of its applicability. Parameter $k$ allows us to vary the scales of time, velocity and viscosity and gives us the possibility to consider any initial and boundary conditions. In general, we discover and classify a rich variety of possibilities and restrictions, which are hidden behind the routine use of the Boussinesq approximation. The paper is devoted to the multiplicity of scalings and related restrictions. We do not study any particular solutions and particular failures of EBA.

\end{abstract}

\section{Introduction \label{sect01}}

In the \emph{equations of Boussinesq approximation} (EBA) the density variations are neglected in the inertial terms of the equations of motion.
EBA are actively employed to describe the flows of a stratified fluid, as well as convective flows. The use of EBA is so common that it is often accepted as a `starting' or `original' system of governing equations in many studies.
 A number of papers and books gives purely physical justification of the EBA (or does not give any justifications at all), see \emph{e.g.} \cite{GershuniC, Craik, Moffatt, Lighthill, YihF, Grimshaw, DrazinStab}.
In this note we present a general analysis of EBA from a viewpoint of the asymptotic theory. The asymptotic approach to EBA has already been exploited in few papers. In \cite{Ver,Zey} the EBA for a compressible fluid are analyzed. The aim of the authors was to `filter out' acoustic waves with the Mach number used as a small parameter. \cite{Long} analysed the EBA for an incompressible fluid, however, he used it alone with the long-wave approximation. The generalization to magneto-hydrodynamic flows has been presented by \cite{Dave}. In our paper we avoid all those very interesting generalizations and concentrate our attention at the simplest case of an incompressible fluid.

We derive EBA  with the use of the same `asymptotic spirit' as, say, in the derivations of KDV equation, Boussinesq equation for surface waves,  Schrodinger equation in various areas of fluid dynamics. Our analysis shows the multiplicity of the scalings and, simultaneously, their internal restrictions.
For example, one such restriction  shows that if the velocity is small (of order $\varepsilon$)
then the time-scale must be of order $1/\varepsilon$.

The aim of this paper is to establish and  analyse the multiplicity of scalings leading to EBA, the quality of EBA models, and their restrictions.
The sensitive point to consider is: the authors, who consider EBA as a `starting' system of governing equations, introduce a scaling required for their particular problems. At the same time EBA require their own scaling, which is not unique but rather restrictive. It is apparent that in a systematic approach these two independently introduced scalings should be incorporated into one scaling, which simultaneously  produces EBA and other desired properties of the equations (long waves, multi-scale, \emph{etc.}). The possibility of such an incorporation should be analysed separately, especially for multi-scale theories.

Our  analysis is concentrated at the general conditions of applicability and quality of EBA with the message that EBA should not be used `blindly', as an original governing equation. In our view, the discovered multiplicity and conditions of applicability rise interesting questions to some existing models and results on nonlinear waves, nonlinear stability, and solitons in a continuously stratified flows, see \cite{Craik, Grimshaw1}. Another area, where our results could be implemented, is vibrational convection, see \cite{GershuniV}. Contrary to \cite{Long}, we do not target the particular solutions of EBA and the failures of EBA.

\section{General scaling and basic equations}

The dimensional equations describing three-dimensional flows of an incompressible stratified viscous fluid are
\begin{eqnarray}
&&\rho^* \va^*=\rho^*[\vu^*_{t^*}+(\vu^*\cdot\nabla^*)\vu^*]=-\nabla^* p^*-\rho^*\nabla^*\Phi^*+\mu^*\Delta^*\vu^*\label{maineq-dim}\\
&& \rho^*_{t^*}+ (\vu^*\cdot\nabla^*)\rho^*=0,\quad \nabla^*\cdot\vu^*=0\label{maineq-dim1}\\
&& \mu^*=\const,\quad\nabla^*\equiv (\partial/\partial x_1^*, \partial/\partial x_2^*, \partial/\partial x_3^*),\quad \Delta^*\equiv\nabla^{*2}\nonumber
\end{eqnarray}
where $t^*$ and $\vx^*=(\vx_1^*,\vx_2^*, \vx_3^*)$ are time and cartesian coordinates;   $\va^*(\vx^*,t^*)$, $\vu^*(\vx^*,t^*)$, $\rho^*(\vx^*,t^*)$, $p^*(\vx^*,t^*)$, $\mu^*$ are the fields of acceleration, velocity, density, pressure, and constant viscosity correspondingly; $\Phi^*(\vx^*)$ is a given potential of external force, such that the acceleration of gravity is $\vg^*=-\nabla^*\Phi^*$;  the subscripts of independent variables denote partial derivatives and asterisks stand for dimensional variables.
The exact solution of these equations is an equilibrium of a homogeneous fluid:
\begin{eqnarray}
 && \rho^*(\vx^*,t^*)=R=\const,\quad \vu^*(\vx^*,t^*)=0,\quad p^*(\vx^*,t^*)=-R\,\Phi^*(\vx^*)+\const\label{mainsoln}
 \end{eqnarray}
 It is also called \emph{a reference solution}. Below we  consider the class of solutions
\begin{eqnarray}
\rho^*=\rho^*(\vx^*,t^*),\quad \vu^*=\vu^*(\vx^*,t^*), \quad p^*=p^*(\vx^*,t^*)\label{genflow}
\end{eqnarray}
which represent small deviations from \eqref{mainsoln}, characterised by two small parameters
\begin{eqnarray}
 && \delta_1\equiv\frac{||\rho^*-R||}{R}\ll 1,\quad \delta_2\equiv\frac{||\va^*||}{||\vg^*||}\sim \frac{||\vu^*_{t^*}||}{||\nabla^*\Phi^*||}\sim \frac{||(\vu^*\cdot\nabla^*)\vu^*||}{||\nabla^*\Phi^*||}\ll 1\label{physics}
\end{eqnarray}
where the particular choice of mathematical norms is not important; for example they can be $C$-norms.
Two key EBA assumptions \eqref{physics} are, as a rule, accepted by all authors, often implicitly.
The list of characteristic parameters for solutions \eqref{genflow} is taken as
\begin{eqnarray}
{L},\quad R,\quad{G}, \quad \delta_1, \quad \delta_2\label{params0}
\end{eqnarray}
where ${L}$ is a characteristic length, ${G}$ is a characteristic value for $|\vg^*|$.
We also take $\mu^*=R\nu^*=const$, where $\nu^*$ is a constant value of dimensional kinematic viscosity.
 Parameter $\nu^*$ is not included into the list \eqref{params0}, since our intention is to consider an inviscid fluid $\nu^*=0$ as a primary case and a viscous fluid $\nu^*\neq 0$ only as an indication of related possibilities. For the latter case $\nu^*$ must be included into the list \eqref{params0}, while the dependent viscous length scale is $L_\nu=(\nu^{*2}/G)^{1/3}$. Then, we consider $L=L_\nu$, in order to avoid a consideration of boundary layers, \emph{etc.}

The imposed conditions \eqref{physics} appear as two independent small parameters $\delta_1$ and $\delta_2$.
They must be expressed as a path $(\delta_1(\varepsilon), \delta_2(\varepsilon))$ in the plane $(\delta_1,\delta_2)$ with the limit $(\delta_1,\delta_2)\to(0,0)$ in order to form a distinguished path (or distinguished limit).
This path must be parameterized with a single small parameter $\varepsilon\to 0$.  Hence, the list \eqref{params0} must be replaced with
\begin{eqnarray}
{L},\quad R,\quad{G}, \quad \varepsilon\label{params}
\end{eqnarray}
A very general \emph{ansatz} of asymptotic analysis is: all independent parameters and functions are allowed to be
scaled with the use of a basic dimensionless parameter, which, in our case, is $\varepsilon$.  It means, that we can introduce, say, dimensionless velocity $\vu$ by an expression $\vu^*=\varepsilon^k{L}^{1/2}{G}^{1/2}\vu$ with an arbitrary real number $k$ (not just $\vu^*={L}^{1/2}{G}^{1/2}\vu$), \emph{etc}.
This kind of scaling is common in the asymptotic analysisand  in perturbation methods, and, in particular, in  fluid dynamics, where, say,  the length-scale for viscous flows can be chosen as $L\sqrt{Re}$ ($Re$ is Reynolds number).
Hence, as a starting point of our asymptotic consideration, we introduce a `flexible' scaling, depending on indefinite real parameters $q,k,m,n,r$
\begin{eqnarray}
&&\vx^*=L\vx\label{scale-x}\\
&&\rho^*=R(1+\varepsilon^q\sigma),\quad q>0\label{scale-rho}\\
&& \vu^*=\varepsilon^k{L}^{1/2}{G}^{1/2}\vu\label{scale-u}\\
&&  t^*= {\varepsilon}^{-k}{L}^{1/2}{G}^{-1/2}\ {t}\label{scale-time}\\
&& \Phi^*=\varepsilon^m{L}\, {G}\, \Phi\label{scale-Phi}\\
&& p^*=R{{L}{G}}(-\varepsilon^m\Phi+\varepsilon^n \pi)\label{scale-p}\\
&& \mu^*=\varepsilon^rR {L}^{3/2}{G}^{1/2}\nu\label{scale-mu}
\end{eqnarray}
where $\nu=\const$. The following \emph{imposed requirements} (IR1-4) are crucial
for the understanding of \eqref{scale-x}-\eqref{scale-mu}:

(IR1) The key restriction $q>0$  provides the smallness of density deviations (from constant density $R$), required by $\delta_1\ll 1$ \eqref{physics}; one can take  $\delta_1=\varepsilon^q$.

(IR2) The introduction of dimensionless time $t$ \eqref{scale-time} with the same parameter $k$ as  in $\vu^*$ \eqref{scale-u} is required for the balance
\begin{eqnarray}
{||\vu^*_{t^*}||}\sim {||(\vu^*\cdot\nabla^*)\vu^*||} \label{nonlin}
\end{eqnarray}
which represents a part of \eqref{physics}. The restriction \eqref{nonlin} can be revoked in the papers which concentrate solely on linearized models, see \cite{Wood}; we do not study this simplified direction here. Hence, the field of acceleration is
$$
\va^*=\varepsilon^{2k} G\va(\vx,t);\quad \va\equiv\vu_t+(\vu\cdot\nabla)\vu
$$

(IR3) The appearance of the same power $m$ in \eqref{scale-Phi} and  \eqref{scale-p} is necessary for the mutual cancelation of a gravity term and a pressure term, describing a hydrostatic equilibrium in \eqref{maineq-dim} and \eqref{mainsoln}.

(IR4) We set all dimensionless functions and parameters $\sigma, \vu, \Phi, \pi, \nu$ in \eqref{scale-rho}-\eqref{scale-mu} of order one
\begin{eqnarray}\label{O1}
&&\sigma(\vx,t)\sim O(1),\ \vu(\vx,t)\sim \vO(1),\ \pi(\vx,t)\sim O(1),\ \Phi(\vx)\sim O(1),\ \nu\sim O(1)
\end{eqnarray}
which means that dimensionless magnitudes of the related fields and parameters are described only by  related powers of $\varepsilon$.
We also accept that all required for our study derivatives of $\sigma, \vu, \Phi, \pi$ \eqref{O1} are of order one.

The substitution of \eqref{scale-x}-\eqref{scale-mu} into \eqref{maineq-dim},\eqref{maineq-dim1}  yields
\begin{eqnarray}
&&(1+\varepsilon^q \sigma)\varepsilon^{2k}[\vu_t+(\vu\cdot\nabla)\vu]=-\varepsilon^n\nabla \pi-\varepsilon^{q+m}\sigma\nabla\Phi+\varepsilon^{k+r}\nu\Delta \vu\label{pre-Buss-0}\\
&& \sigma_t+(\vu\cdot\nabla)\sigma=0,\quad \nabla\cdot\vu=0\nonumber
\end{eqnarray}
where the required (for the obtaining of EBA) setting is to choose all involved terms of the same order in $\varepsilon$:
\begin{eqnarray}
2k=n=q+m=k+r\label{relation}
\end{eqnarray}
It leads to the equations
\begin{eqnarray}
&&(1+\varepsilon^q \sigma)[\vu_t+(\vu\cdot\nabla)\vu]=-\nabla \pi-\sigma\nabla\Phi+\nu\Delta \vu\label{pre-Buss-1}\\
&& \sigma_t+(\vu\cdot\nabla)\sigma=0,\quad \nabla\cdot\vu=0\nonumber
\end{eqnarray}
which below are referred to as the \emph{main equations}, which explicitly contains a small parameter $\delta\equiv \varepsilon^q$.
Since the relations \eqref{relation} give us three connections between five parameters, we rewrite \eqref{scale-x}-\eqref{scale-mu} with the use of two parameters $q$ and $k$ only:
\begin{eqnarray}
&&\vx^*=L\vx\label{scale-x-1}\\
&&\rho^*=R(1+\varepsilon^q\sigma),\quad q>0\label{scale-rho-1}\\
&& \vu^*=\varepsilon^k{L}^{1/2}{G}^{1/2}\vu\label{scale-u-1}\\
&&  t^*= {\varepsilon}^{-k}{L}^{1/2}{G}^{-1/2}\ {t}\label{scale-time-1}\\
&& \Phi^*=\varepsilon^{2k-q}{L}\, {G}\, \Phi\label{scale-Phi-1}\\
&& p^*=R{{L}{G}}(-\varepsilon^{2k-q}\Phi+\varepsilon^{2k} \pi)\label{scale-p-1}\\
&& \mu^*=\varepsilon^kR {L}^{3/2}{G}^{1/2}\nu\label{scale-mu-1}
\end{eqnarray}
The \emph{observations} O1-3 are:

(O1) Equations \eqref{pre-Buss-1}, taken for an arbitrary value of $\varepsilon$, still give us the exact governing equations describing incompressible stratified flows. However, for  $\varepsilon\to 0$ these equations describe only some special classes of stratified flows.
Taken together, equations \eqref{pre-Buss-1}-\eqref{scale-mu-1} give us an infinite number of asymptotic models of  incompressible stratified flows, one model for each pair $(q,k)$.

(O2) We notice, that the dimensionless equations of motion \eqref{pre-Buss-1} contain only parameter $q$, while the underlining scaling \eqref{scale-x-1}-\eqref{scale-mu-1} possesses an additional freedom given by two parameters $q$ and $k$. As soon as $q$ is chosen, the restriction on the choice of $k$ is in related physics: we have to decide what is the asymptotic behaviour of $\Phi^*\sim\varepsilon^{2k-q}$ at $\varepsilon\to 0$. It is `singular' for ${2k-q}<0$ however it does not prevent us against of using it (see Discussion).

(O3) For any $k$, the amplitudes of velocity and viscosity are of the order $\varepsilon^k$. It brings us to an expected conclusion: the Reynolds numbers ($Re$) are of order one in all viscous flows. Indeed, in \eqref{relation} we have accepted that a viscous term is of the same order as the other terms in the Navier-Stokes equations, which corresponds to $Re\sim 1$. The consideration of high or low $Re$ will take us to different asymptotic models, which we avoid  in this paper.

\section{Three levels of Boussinesq approximations}

 Due to $q>0$, the main equation \eqref{pre-Buss-1} explicitly contains a small parameter; hence its solutions for  $\varepsilon\to 0$ can be considered as  series in $\varepsilon$:
\begin{eqnarray}\label{series1}
(\sigma,\vu,\pi)=\sum_{\alpha=0}^\infty \varepsilon^\alpha (\sigma_\alpha,\vu_\alpha,\pi_\alpha),\quad \alpha=0,1,2,\dots
\end{eqnarray}
The substitution of \eqref{series1} into \eqref{pre-Buss-1} produces the equations of successive approximations in $\varepsilon$.
The use of integers $\alpha$ means that we should operate only with the integers $q=1,2,\dots$ and $k=0, \pm1,\pm2,\dots$.
The leading approximation ($\alpha=0$) is
\begin{eqnarray}
&&\vu_{0t}+(\vu_0\cdot\nabla)\vu_0=-\nabla \pi_0-\rho_0\nabla\Phi+\nu\Delta \vu_0\label{pre-Buss-1-0}\\
&& \rho_{0t}+(\vu_0\cdot\nabla)\rho_0=0,\quad \nabla\cdot\vu_0=0\nonumber
\end{eqnarray}
which can be immediately recognized as \emph{the equations of Boussinesq approximation} (EBA) for unknown fields $\rho_0,\vu_0,\pi_0$.
To emphasise the use of EBA as the `governing' equations of stratified flows, we change the notations in \eqref{pre-Buss-1-0} as
$\sigma_0\mapsto \rho,\ \vu_0\mapsto\vv,\ \pi_0\mapsto p$. That produces a conventional dimensionless form of  EBA:
\begin{eqnarray}
&&\vv_{t}+(\vv\cdot\nabla)\vv=-\nabla p-\rho\nabla\Phi+\nu\Delta\vv\label{EBA}\\
&& \rho_{t}+(\vv\cdot\nabla)\rho=0,\quad \nabla\cdot\vv=0\nonumber
\end{eqnarray}
Their solutions also can be presented as the amplitude series
\begin{eqnarray}\label{seriesB}
(\rho,\vv,p)=\sum_{\beta=0}^\infty \varepsilon^\beta (\rho_\beta,\vv_\beta,p_\beta),\quad \beta=0,1,2,\dots
\end{eqnarray}
where we have accepted, that the amplitudes are described by the same small parameter $\varepsilon$. Here we should explain, why the amplitude
parameter for the main equations \eqref{series1} and EBA \eqref{seriesB} is the same $\varepsilon$.
Indeed, it is possible to introduce different  amplitude parameters, however such a complication can be seen as an introduction of two different small parameters for the same physical entity (for the amplitude of a solution).

The linearization of \eqref{EBA} (which corresponds to the term $\beta=1$ in \eqref{seriesB}) is
\begin{eqnarray}
&&\vv_{1t}+(\vv_0\cdot\nabla)\vv_1+(\vv_1\cdot\nabla)\vv_0=-\nabla p_1-\rho_1\nabla\Phi+\nu\Delta\vv_1\label{EBA-lin}\\
&& \rho_{1t}+(\vv_0\cdot\nabla)\rho_1+(\vv_1\cdot\nabla)\rho_0=0,\quad \nabla\cdot\vv_1=0\nonumber
\end{eqnarray}
At the same time, the first approximation (which corresponds to the term $\alpha=1$ in \eqref{series1}) of the main equations \eqref{pre-Buss-1} (written for the lowest value $q=1$), is:
\begin{eqnarray}
&&\vu_{1t}+(\vu_0\cdot\nabla)\vu_1+(\vu_1\cdot\nabla)\vu_0 +\vA_0=-\nabla \pi_1-\sigma_1\nabla\Phi+\vu\Delta\vu_1\label{pre-Buss-1-1}\\
&& \sigma_{1t}+(\vu_0\cdot\nabla)\sigma_1+(\vu_1\cdot\nabla)\sigma_0=0,\quad \nabla\cdot\vu_1=0\nonumber
\end{eqnarray}
 Comparison of \eqref{EBA-lin} and \eqref{pre-Buss-1-1} shows that
 \begin{eqnarray}
\vA_0\equiv \rho_0 \va_0=\rho_0 [\vu_{0t}+(\vu_{0}\cdot\nabla)\vu_{0}]\label{spoil}
\end{eqnarray}
represents a `spoiling term', making the linearized version \eqref{pre-Buss-1-1} of the main equations \eqref{pre-Buss-1} different from the linearized version of  EBA \eqref{EBA-lin}.
If we take $q= 2$ or above, then this `spoiling term' does not appear in the equations corresponding to $\alpha=1$; for the first time it appears in the approximation of the order $\alpha=q$. Hence, we introduce \emph{three different levels for the quality of approximations of the main equations  given by  EBA:}
\begin{eqnarray}
&&\text{\underline{Level 1}}:\quad  q=1\quad \text{\texttt{poor approximation}}\label{levels}\\
&&\text{\underline{Level 2}}:\quad q=2\quad \nonumber \text{\texttt{reasonable approximation}}\\
&&\text{\underline{Level 3}}:\quad q\geq 3\quad \text{\texttt{good approximation}} \nonumber
\end{eqnarray}
For a poor approximation (Level 1) only  EBA itself \eqref{EBA} coincides with the leading term $\alpha=0$ of the main equation, while the linearized versions $\alpha=1$ and $\beta=1$ are different from each other.
For a reasonable approximation (Level 2) we have the coinciding of  EBA and its linearization $\beta=1$ with the approximations $\alpha=0$ and $\alpha=1$ of the main equations. Physically, it guarantees the same results for linear waves and linear stability for the EBA and for the main equations.
For a good approximation (Level 3)  the series for EBA coincides with that for the main equations up to $\alpha=2$. Physically, this means that both linear perturbations and  weakly nonlinear perturbations for EBA and for the main equations are the same.

\section{Classification of the EBA models.}

Asymptotic models with $k=0$, $k>0$, and $k<0$ are qualitatively different,  hence we consider them separately.
For \underline{the models with $k=0$} equations \eqref{scale-x-1}-\eqref{scale-mu-1} yield:
\begin{eqnarray}
&&\vx^*=L\vx\label{scale-x-10}\\
&&\rho^*=R(1+\varepsilon^q\sigma),\quad q>0\label{scale-rho-10}\\
&& \vu^*={L}^{1/2}{G}^{1/2}\vu\label{scale-u-10}\\
&&  t^*= {L}^{1/2}{G}^{-1/2}\ {t}\label{scale-time-10}\\
&& \Phi^*=\varepsilon^{-q}{L}\, {G}\, \Phi\label{scale-Phi-10}\\
&& p^*=R{{L}{G}}(-\varepsilon^{-q}\Phi+ \pi)\label{scale-p-10}\\
&& \mu^*=R {L}^{3/2}{G}^{1/2}\nu\label{scale-mu-10}
\end{eqnarray}
One can see, that an advantage of this model is the  finite limits (at $\varepsilon\to 0$) for $\vu^*$, $t^*$, $\mu^*$, and for the deviation $RLG\pi$ of pressure from a hydrostatic reference state. At the same time, the limit for $\Phi^*$ becomes more `singular' with the increase of  $q$  from Level 1 to Level 3. In other words, these models correspond to `order one' velocity and viscosity and to `usual physical time' in the case of strong gravity. It is important to notice, that a `singular' limit for gravity is perfectly permitted and actively used, see Discussion.


The \underline{models with $k>0$} are characterised by  `slow time' $t^*$, small velocity $\vu^*$ and small viscosity $\nu^*$, which appear in \eqref{scale-x-1}-\eqref{scale-mu-1} for $k>0$.
The gravity is weak for $2k-q>0$ and `strong' for $2k-q<0$.

\underline{A physically  attractive model}, giving a reasonable approximation \eqref{levels},  corresponds to $k=1$ and $q=2$
\begin{eqnarray}
&&\vx^*=L\vx\label{scale-x-main}\\
&&\rho^*=R(1+\varepsilon^2\sigma)\label{scale-rho-main}\\
&& \vu^*=\varepsilon\sqrt{{L}{G}}\vu\label{uscale}\\
&&  t^*= \frac{1}{\varepsilon}\sqrt{{L}/{G}}\ {t}\label{scale-time-main}\\
&& \Phi^*={L}\, {G}\, \Phi\label{scale-Phi-main}\\
&& p^*=R{{L}{G}}(-\Phi+\varepsilon^{2} \pi)\label{scale-p-main}\\
&& \mu^*=\varepsilon R {L}^{3/2}{G}^{1/2}\mu\label{scale-mu-main}
\end{eqnarray}
when the main equations \eqref{pre-Buss-1} are:
\begin{eqnarray}
&&(1+\varepsilon^2 \sigma)[\vu_t+(\vu\cdot\nabla)\vu]=-\nabla \pi-\sigma\nabla\Phi+\mu\Delta \vu\label{pre-Buss-main}\\
&& \sigma_t+(\vu\cdot\nabla)\sigma=0,\quad \nabla\cdot\vu=0\nonumber
\end{eqnarray}
This case is characterised by:
(i) a small amplitude of velocity $|\vu^*|\sim\varepsilon$;
(ii) small (of order $\varepsilon^2$) deviations of density and pressure from  hydrostatic (reference) density and pressure;
(iii) a finite limit for gravity $\Phi^*\sim\varepsilon^0$;
(iv) a long time-scale $t^*\sim 1/\varepsilon$ (or a low characteristic frequency $\sim \varepsilon$);
(v)  small viscosity $\mu^*\sim \varepsilon$.

One may say that  properties (i) and (iii) can be seen as physical advantages. Indeed, the smallness of velocity $\vu^*\sim\varepsilon$ is natural for the waves of small amplitude  or for generic small perturbations; such a smallness is accepted in the majority of related publications. The gravity $\Phi^*\sim\varepsilon^0$ also may be considered as  `physically natural'. Some authors consider only this scaling, which is always written differently due to the presence of additional small parameters, see \emph{e.g.} \cite{Long}). These additional parameters are related to, say, the long-wave approximation or small compressibility.

Finally, \underline{the models with $k<0$} are characterised by the `fast time' variable and the large amplitude velocity and viscosity, see \eqref{scale-x-1}-\eqref{scale-mu-1} for $k<0$.
The gravity is always strong. This class of models has been never studied or used, however the possible application could be to rapidly oscillating flows or to turbulence.

\section{How to choose parameters $q$ and $k$?}

As one have seen the main equations \eqref{pre-Buss-1} contain only  parameter $q=1,2,3,\dots$, while the underlining scaling \eqref{scale-x-1}-\eqref{scale-mu-1} is defined by two parameters $q$ and $k=\pm 1,\pm 2,\pm 3,\dots$.
In  particular problems, these parameters can only appear from the prescribed boundary conditions and initial conditions.

\underline{The role of initial conditions is:} since in \eqref{scale-x-1} we do not re-scale the spatial variable $\vx$, the dimensionless initial conditions can be always written as
\begin{eqnarray}
\rho(\vx,0)=\rho^{\dag}(\vx),\quad \vu(\vx,0)=\vu^{\dag}(\vx)\label{IC}
\end{eqnarray}
with given dimensionless functions $\rho^{\dag}(\vx)$ and $\vu^{\dag}(\vx)$.  Hence, here we must accept that the magnitudes of the solution \eqref{scale-rho-1}, \eqref{scale-u-1} coincide with that of the initial conditions \eqref{IC}.

\underline{The role of boundary conditions} is slightly different for  time-independent and time-dependent cases.
The former case requires the same orders of functions \eqref{scale-rho-1}, \eqref{scale-u-1}, while the latter case additionally requires the same orders in the time-scale \eqref{scale-time-1}.

Summarising, we write that the introduced multiplicity of scalings allows us to consider any amplitudes of the functions and variables, as they appear in the initial and boundary conditions.

An interesting question arises from the particular appearance of $\varepsilon$ in the main equations \eqref{pre-Buss-1}, which contain only a small parameter $\delta\equiv\varepsilon^q$. It suggests that  solutions can be represented as  power series of either $\delta$ or $\varepsilon$. In the former case \eqref{pre-Buss-1} and \eqref{series1} should be replaced with
\begin{eqnarray}
&&(1+\delta\, \rho)[\vu_t+(\vu\cdot\nabla)\vu]=-\nabla p-\rho\nabla\Phi\label{pre-BussD}\\
&& \rho_t+(\vu\cdot\nabla)\rho=0,\quad \nabla\cdot\vu=0\nonumber
\end{eqnarray}\vskip -5mm
\begin{eqnarray}\label{seriesD}
(\rho,\vu,p)=\sum_{\gamma=0}^\infty \delta^\gamma (\rho_\gamma,\vu_\gamma,p_\gamma),\quad \gamma=0,1,2,\dots
\end{eqnarray}
which corresponds to the poor approximation \eqref{levels}, with $q=1$. However, one can also use parameter $\varepsilon$, take $q=2$ or $q\geq 3$ and obtain the reasonable or good approximations. The cost of this improvement will be the increasing of a small parameter. Indeed, if in \eqref{pre-BussD} (for a poor approximation) we choose, say $\delta=10^{-3}$, then for a good approximation with $\delta=\varepsilon^3$ we obtain
$\varepsilon=10^{-1}$. Of course, the increasing of a small parameter makes the converging of series worse. Hence, \emph{we have a counterbalance}: the improving of  the model quality makes the convergence worse and for a lower model quality we improve the convergence of series. The optimization of this counterbalance represents an interesting practical question.
At the same time we notice here that the conversion from $\varepsilon$ to $\delta$ is impossible if $\varepsilon$ appears explicitly in  boundary conditions or in  initial conditions, through \eqref{scale-x-1}-\eqref{scale-mu-1}.

Another restriction on the values of $q$ and $k$ may appear from physics: for a particular application one may `prefer'  strong gravity or weak gravity.
For example, if the value of $k$  (enforced, say, by boundary conditions) is $k=0$, then \eqref{scale-Phi-1} immediately gives us $\Phi^*\sim \varepsilon^{-q}$, hence, with the increase of the quality of the EBA model we have to consider the increasingly strong gravity field.
On the other hand, if we wish to consider $\Phi^*\sim \varepsilon^{0}$ then we have to take $2k=q$, and the option to choose is $k=1, q=2$ \eqref{scale-x-main}-\eqref{scale-mu-main}  with slow time \eqref{scale-time-main}. Other options for $t^*$ will require the change of the magnitude of a gravity field.

 The general conclusion of this section is: a flexible choice of $q$ and $k$ allows to consider a broad variety of problems with different orders of unknown functions and scales involved; simultaneously one can choose the asymptotic models of different (desired) quality.

\section{Discussion}

 (D1) It is surprising, that we were unable to find the scaling \eqref{scale-x-1}-\eqref{scale-mu-1} and even the main equations \eqref{pre-Buss-1} in the literature on stratified flows.
 
 (D2)A question could be raised: what is the relation between the internal time scales (say, the inverse Brunt-Vaisala frequency) and that of \eqref{scale-time-1}? The answer is: after the internal time-scale is properly scaled, it coincides with \eqref{scale-time-1}.

(D3) Another question could be asked: why the main equations \eqref{pre-Buss-1} are `worse' or `more difficult to solve' than the EBA \eqref{EBA}? We do believe that the direct solving of the main equation \eqref{pre-Buss-1} is preferable for many studies.

 (D4) The asymptotic for gravity $\Phi^*\sim\varepsilon^{2k-q}$ at $\varepsilon\to 0$ in \eqref{scale-Phi-1}  is `singular' for ${2k-q}<0$ however it does not prevent one against of using it. For example,  the `singular' asymptotic $\Phi^*\sim 1/\varepsilon$ has been actively used in the vibrational convection, see \cite{Zenkovskaya, Simonenko, GershuniV, Levenshtam}. It is apparent that, without the use of this `singular' asymptotic, the theory of vibrational convection can not be constructed. The physical meaning of the `singular' asymptotic is: we deal with a strong (not infinite) gravity, since in applications $\varepsilon$ is always finite.

(D5) The separation of the EBA  models into  poor, reasonable, and good approximations requires further clarification. 
The particular impacts of `spoiling terms'  \eqref{spoil} should be investigated separately. The target here could be the obtaining of an estimation which could be written as an upper bound for a norm
\begin{eqnarray}\label{close}
||\vu-\vv,\rho-\sigma||\leq C(\varepsilon)||\sigma\va, \vu-\vv,\rho-\sigma||_{t=0}
\end{eqnarray}
where  constant $C$ depends on $\varepsilon$ such that $C\to 0$ when $\varepsilon\to 0$. The presence of $\rho\va$ in \eqref{close} represents an assumption  that the difference between solutions is caused by the `spoiling term', as well as by the differences in initial conditions.
It is especially interesting to compile a list of examples, where particular solutions  of EBA are qualitatively different from the related solutions of the main equations  (taken for small $\varepsilon$). One such example is presented in \cite{Long}.

(D6) Particular solutions, illustrating the results of this paper,  must be essentially unsteady, since the steady versions of EBA and the exact equations for steady flows are mathematically equivalent to each other, see \cite{YihS}.
 The most useful for our studies would be an exact unsteady solution of EBA and the main equations, satisfying the same initial and boundary conditions.
 However, the unsteady exact solutions for stratified flows  are almost unknown, a noticeable exception is \cite{CraikExact}.

(D7) The presented derivations raise some interesting questions about the scalings used in the studies of nonlinear waves (including solitons) and nonlinear stability of shear flows in the presence of density stratification, such as considered in \cite{Craik, Grimshaw1}. It would be interesting to incorporate the presented scaling required for EBA to the elaborated scalings of the related studies. The combining of two scalings, which have been introduced independently from each other, may bring some surprises.

(D8) Interesting developments can appear in  various cases where the Boussinesq approximation is used simultaneously with additional independent small parameters (for example, long waves or an inverse frequency of externally imposed oscillations).
It is likely that some developments (and even some interesting contradictions between different requirements to scalings used for different small parameters) may appear in the problem of vibrational convection and in the problem of stability and waves in a vibrating stratified fluid \cite{GershuniV, Yudovich, Zenkovskaya}, where two different time scales are employed from very beginning.

(D9) The `filtering out' of acoustic oscillations was performed in \cite{Ver, Zey}. An alternative to this studies approach could be the  deriving of averaged equations with the use of two-timing averaging over the high-frequency oscillations, as it has been done in \cite{VladIlin, VladimirovL, VladProc}.

\begin{acknowledgments}
The authors would like to express special gratitude to Profs. A.D.D. Craik, R.H.J. Grimshaw, and H.K. Moffatt for reading this manuscript and making useful critical comments. Many thanks to Profs. I.A. Eltayeb, D.W. Hughes, A.R. Kacimov, M.R.E. Proctor, and M.M. Rahman for helpful discussions.
This research is supported by the grant IG/SCI/ DOMS/16/13 from Sultan Qaboos University, Oman.
\end{acknowledgments}

\end{document}